\documentclass[aps,pra,twocolumn]{revtex4}

\usepackage{epsfig,pstricks}

\usepackage{amsmath,amssymb}
\usepackage{color}

\DeclareMathOperator{\sign}{sign}

\def\CC{{\rm\kern.24em \vrule width.04em height1.46ex depth-.07ex
\kern-.30em C}}
\def\P{{\rm I\kern-.25em P}}
\def\NN{{\rm I\kern-.25em N}}
\def\RR{{\rm
         \vrule width.04em height1.58ex depth-.0ex
         \kern-.04em R}}
\def\id{{\rm 1\kern-.22em l}}
\def\ZZ{{\sf Z\kern-.44em Z}}

\newtheorem{pdef}{Definition}[section]

\newenvironment{eqblock}[2]{\beq\label{#2}\begin{array}{#1}}{\end{array}
                                \eeq}
\newenvironment{neqblock}[1]{\[\begin{array}{#1}}{\end{array}\]}
\newcommand{\braket}[2]{\langle #1 | #2 \rangle}
\newcommand{\ketbra}[1]{\ensuremath{| #1 \rangle \langle #1 |}}
\newcommand{\fat}[1]{\mbox{\boldmath $ #1 $\unboldmath}}
\newcommand{\beqb}{\begin{eqblock}}
\newcommand{\eeqb}{\end{eqblock}} 
\newcommand{\nbeqb}{\begin{neqblock}}
\newcommand{\neeqb}{\end{neqblock}} 

\newcommand{\beq}{\begin{equation}}
\newcommand{\beqa}{\begin{eqnarray}}
\newcommand{\eeq}{\end{equation}}
\newcommand{\eeqa}{\end{eqnarray}}
\newcommand{\nbeqa}{\begin{eqnarray*}}
\newcommand{\neeqa}{\end{eqnarray*}}

\newcommand{\ket}[1]{| #1 \rangle}

\def\DJo{$\;$\kern-.4em \hbox{D\kern-.8em\raise.15ex\hbox{--}\kern.35em okovi\'c}}

\begin{document}

\title{Exact zeros of entanglement for arbitrary rank-two mixtures: how a geometric view of the zero-polytope makes life more easy}
\author{Andreas Osterloh}
\affiliation{Institut f\"ur Theoretische Physik, 
         Universit\"at Duisburg-Essen, D-47048 Duisburg, Germany.}
\email{andreas.osterloh@uni-due.de}
\begin{abstract}
Here I present a method how intersections of a certain density matrix of rank two with 
the zero-polytope can be calculated exactly. This is a purely geometrical procedure which
thereby is applicable to obtaining the zeros of SL- and SU-invariant entanglement measures
of arbitrary polynomial degree. 
I explain this method in detail for a recently unsolved problem.
In particular, I show how a three-dimensional view,
namely in terms of the Boch-sphere analogy, solves this problem immediately. 
To this end, I determine the zero-polytope of the three-tangle, 
which is an exact result up to computer accuracy,
and calculate upper bounds to its convex roof which are below the linearized upper bound. 
The zeros of the three-tangle (in this case) induced by the zero-polytope (zero-simplex)
are exact values. 
I apply this procedure to a superposition of the four qubit GHZ- 
and W-state. It can however be applied to every case one has under consideration, including an arbitrary 
polynomial convex-roof measure of entanglement and for arbitrary local dimension. 

\end{abstract}

\maketitle

\section{Introduction}

Entanglement is one of the key features of quantum mechanics
that is omnipresent in mutually interacting systems.
Measures of entanglement are minimally invariant under local unitaries\cite{MONOTONES}.
This invariance emerges when dealing with the concept of Local Operations combined with 
Classical Communication (LOCC).
It has however soon been realized that 
this invarianz group has to be extended to the special
linear group\cite{Duer00,VerstraeteDM02,VerstraeteDMV02}
since in general Stochastic Local Oparations combined with 
Classical Communication (SLOCC) have to be included.
Thus, a state $\ket{\psi}$ is said to be equivalent to the state 
$\ket{\psi'}:=(A_1\otimes\dots\otimes A_q)\ket{\psi}$ for $A_i\in SL(d_i)$, and
for each SL-invariant measure $\tau$ of entanglement we have
$\tau(\ket{\psi})=\tau(\ket{\psi'})$.
Every such SL-invariant entanglement measure can be decomposed into polynomial 
measures of entanglement of homogeneous degree.\\
The entanglement content of a mixed state is represented by the convex-roof expression of
the entanglement measure of interest\cite{MONOTONES}. 
Whereas it is more easy to write the convex-roof down 
than to really calculate it, it has shown to be an exactly solvable task for measures,
which are SL-invariant homogeneous polynomials of rank two, 
as the concurrence\cite{Wootters98,Uhlmann00}, respectively convex functions of them.
In this simple case, the optimal decomposition has a continuous degeneracy, 
which is a key ingredient to the exact solution. 
However, already if the homogeneous degree is four, this degeneracy is lost in general and
one is left with a typically unique solution in terms of normalized states, 
not considering global phases and permutations of the states. 
It has therefore become one of the central problems in modern physics
to 'tame' the convex-roof\cite{Jungnitsch11}.
First steps into this direction have been gone in Refs.~\cite{LOSU,EOSU,KENNLINIE} 
where lower bounds for rank two density matrices have been addressed
with some thoughts about the more general case\cite{KENNLINIE}. In some specific cases
this lower bound coincides with the convex-roof solution. With these solutions,
certain particular cases
for rank three density matrices\cite{Jung09} and even higher rank\cite{HigherRankTau3},
which are all constructions out of separable states, have followed.

The convexified minimal {\em characteristic curve}~\cite{LOSU,EOSU,KENNLINIE} 
of the entanglement measure under consideration has been singled out as a lower bound to 
any possible decomposition of $\rho$.
This has been advanced to calculate lower bounds to the three-tangle 
of density matrices with general rank\cite{Eltschka2012,Siewert2012,Eltschka2014},
a lower bound which was shown to be
sharp for the class of states with the symmetry of the GHZ-state, termed {\em GHZ-symmetry}.
This technique for obtaining lower bounds has served later for demonstrating bound entanglement with 
positive partial transpose for qutrit states\cite{SiewertPPT2Qutrits}.\\
In the meantime several algorithms providing with upper bounds 
emerged\cite{CaoZhouGuoHe10,Rodriquez14,OstUpperBound16}, 
where Ref.~\cite{OstUpperBound16} is departing from the solution for the zero-polytope for
rank-two density matrices.
However, also applications of the original method provided in \cite{LOSU,EOSU,KENNLINIE} are
still challenging\cite{OstNineClasses16,JungPark16}.
In their recent contribution, Jung and Park have tempted to test the monogamy relations of 
Coffman, Kundu and Wootters (CKW)\cite{Coffman00}
and for the negativity\cite{Ou07,HeVidal15} towards possible 
extended versions\cite{AdessoRegula14,AdessoRegula14b,AdessoRegula16,Karmakar16}.
They succeded for the negativity, however they encountered problems for the 
Coffman-Kundu-Wootters-monogamy, which they highlighted using a toy-example in their appendix.
The main difference to the case depicted in Ref.~\cite{LOSU} was
that no three zeros of the three-tangle coincided for a given probability $p\in [0,1]$.
Hence their characteristic curves had zeros at three different probabilities.
There, the case of non-coinciding zeros of the characteristic curves was posed as an open problem.\\
We first focus on their toy-example
since it shows 1) how using $C_3:=\sqrt{|\tau_3|}$ instead of $|\tau_3|$ can help 
in calculating meaningful upper bounds of its convex-roof, and 2) the impact not coinciding roots
have onto the three-tangle of the state under consideration. 
The intervals where the mixed three-tangle is zero
can be obtained in a simple geometrical way: they are numerically exact results.

This work is outlined as follows: in the next section I briefly focus on the method and 
give as an example the three-tangle as SL-invariant homogeneous polynomial of degree 4
with reference to \cite{JungPark16}.
Next, I apply this method to the toy-example of Ref.~\cite{JungPark16} in section~\ref{JungPark}
and come to some general states in section~\ref{general}. I briefly comment on extended monogamy relations in section~\ref{monogamy} before making concluding remarks in section~\ref{concls}.

\section{Preliminaries}\label{prelim}

Measures of entanglement are minimally invariant under local unitaries
$\prod_{i=1}^{q;\otimes}SU(d_i)$\cite{MONOTONES}
where $q$ is the number of local objects of dimension $d_i\, ,\ i=1,\dots,q$ 
which are beeing considered. Hence, all states 
$\ket{\psi'}:=(U_1\otimes\dots\otimes U_q)\ket{\psi}$ with $U_i\in SU(d_i)$ are 
considered equivalent. An SU-invariant measure of entanglement ${\cal M}$
satisfies 
\beq
{\cal M}(\ket{\psi'})={\cal M}(\ket{\psi})\; .
\eeq
This invariance is connected to Local Operations combined with 
Classical Communication (LOCC).
It has however been realized that 
this invarianz group has to be extended to the special
linear version $\prod_{i=1}^{q;\otimes}SL(d_i)$\cite{Duer00,VerstraeteDM02,VerstraeteDMV02}
since in general Stochastic Local Oparations combined with 
Classical Communication (SLOCC) must be considered.
There, a state $\ket{\psi}$ is said to be equivalent to the state 
$\ket{\psi'}:=(A_1\otimes\dots\otimes A_q)\ket{\psi}$ for $A_i\in SL(d_i)$, and
for each SL-invariant measure $\tau$ of entanglement holds
\beq
\tau(\ket{\psi})=\tau(\ket{\psi'})\; .
\eeq 
Every SL-invariant entanglement measure can be decomposed into polynomial 
measures of entanglement of homogeneous degree.
I will for brevity write $\tau(\psi)$ for $\tau(\ket{\psi})$. 

It is however remarked that the entanglement content in the 
state is nevertheless modified in that the modulus $\braket{\psi}{\psi}$ is 
modified in general by SL-operations in contrast to the SU-invariance.

I will consider $C_3:=\sqrt{|\tau_3|}$ as entanglement measures, where
the threetangle $|\tau_3|$ has been defined as\cite{Coffman00} 
(see also in Refs.~\cite{Wong00,VerstraeteDM03,OS04})
\nbeqa
\tau_3 &=& d_1 - 2d_2 + 4d_3  \\
  d_1&=& \psi^2_{000}\psi^2_{111} + \psi^2_{001}\psi^2_{110} + \psi^2_{010}\psi^2_{101}+ \psi^2_{100}\psi^2_{011} \\
  d_2&=& \psi_{000}\psi_{111}\psi_{011}\psi_{100} + \psi_{000}\psi_{111}\psi_{101}\psi_{010}\\ 
    &&+ \psi_{000}\psi_{111}\psi_{110}\psi_{001} + \psi_{011}\psi_{100}\psi_{101}\psi_{010}\\
    &&+ \psi_{011}\psi_{100}\psi_{110}\psi_{001} + \psi_{101}\psi_{010}\psi_{110}\psi_{001}\\
  d_3&=& \psi_{000}\psi_{110}\psi_{101}\psi_{011} + \psi_{111}\psi_{001}\psi_{010}\psi_{100}\ \ ,
\neeqa
and coincides with the three-qubit hyperdeterminant\cite{Cayley,Miyake02}. 
It is the only continuous SL-invariant here, 
meaning that every other such SL-invariant for three qubits can be expressed 
as a function of $\tau_3$.

\begin{figure}
\centering
\includegraphics[viewport= 190 570 440 800,clip,width=.9\linewidth]{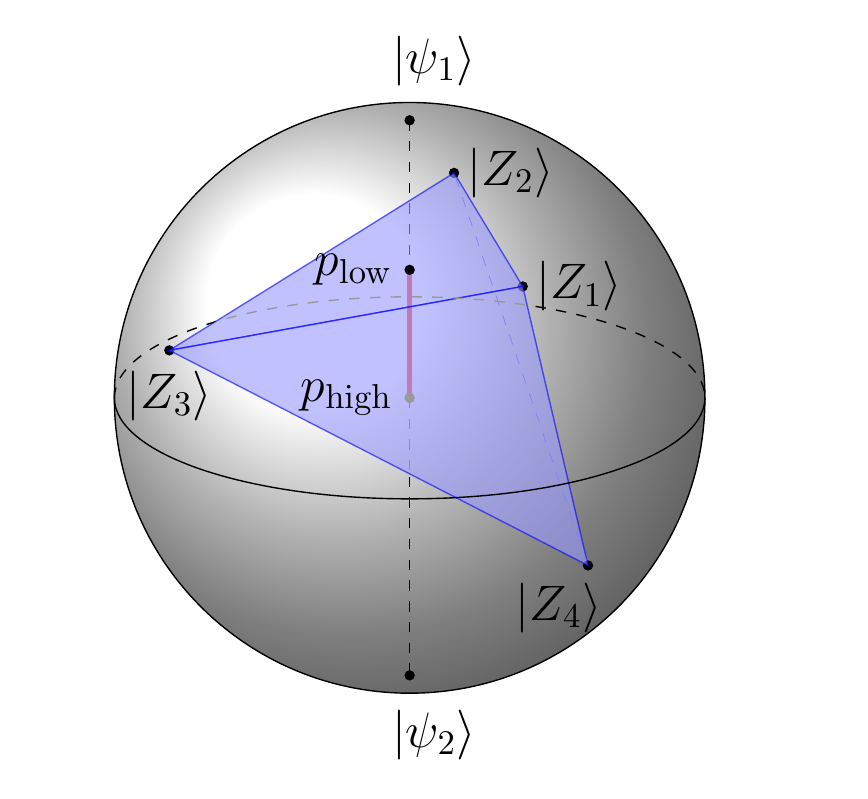}
\caption{An example for a (homogeneous) polynomial SL-invariant $\tau$ for a density matrix of 
rank two, $\rho(p)=p \ketbra{\psi_1} + (1-p) \ketbra{\psi_2}$ is drawn in the Bloch sphere picture. 
The polynomial invariant has the four solutions $\ket{Z_i}$ for $i\in \{1,\dots,4\}$ defining
the zero-polytope. The intersection of this polytope with the line connecting 
$\ket{\psi_1}$ and $\ket{\psi_2}$ 
leads to an interval $[p_{\rm low},p_{\rm high}]$ of vanishing $\tau[\rho(p)]$.
When this intersection is empty, this means that $\rho(p)$ is always entangled as measured by $\tau$.}
\label{Bloch}
\end{figure}

\section{Geometric view of the zero-polytope}

For rank two density matrices $\rho$, the states in the range of $\rho$
can be written as
\beq
\ket{\psi(z)}:=\ket{\psi_1}+z \ket{\psi_2}\; ,
\eeq
with eigenstates $\ket{\psi_i}$ of $\rho$, and $z\in\CC$\cite{KENNLINIE}.
An entanglement measure $\tau$ vanishes precisely on the polytope with the states 
$\ket{\psi(z_0)}$ as vertices, where $z_0\in\CC$ satisfies the equation $\tau(\psi(z_0))=0$; 
this object is called the {\em zero-polytope}\cite{LOSU,KENNLINIE} 
(see also Ref.~\cite{OstNineClasses16}).
One can hence check what triangle between vertices of the zero-polytope has an intersection with the
line connecting $\ket{\psi_1}$ with $\ket{\psi_2}$
at some $p_{0;i}$ for $i\in {\cal I}$.
The values $p_{low}=\min_{i\in {\cal I}} p_{0;i}$ and $p_{high}=\max_{i\in {\cal I}}p_{0;i}$
is the interval where $\rho(p):=p \ketbra{\psi_1} + (1-p)\ketbra{\psi_2}$ is zero.
I have used here a part of the algorithm described in Ref.~\cite{OstUpperBound16} 
(see Eqs.~(10) and (11) therein).
This procedure is illustrated in Fig.~\ref{Bloch} 
where I give an example for a polynomial of (homogeneous) degree four on the Bloch sphere.

For density matrices of higher rank $R$, the states in the range of $\rho$ can be written as
\beq
\ket{\psi(z_1,\dots,z_{R-1})}:=\ket{\psi_1}+z_1 \ket{\psi_2}+ \dots + z_{R-1} \ket{\psi_{R}}\; ,
\eeq
and the zero-polytope turns into the convexification of
the {\em zero-manifold} made out of all the solutions of $\tau(\psi(z_{0;1},\dots,z_{0;R-1}))=0$.

\section{The toy example raised by Jung and Park}\label{JungPark}

To show this method at work, I choose the toy-example out of the appendix of Ref.~\cite{JungPark16}.
\subsection{The geometric view}\label{geometric:0}
\begin{figure}
\centering
\includegraphics[width=.9\linewidth]{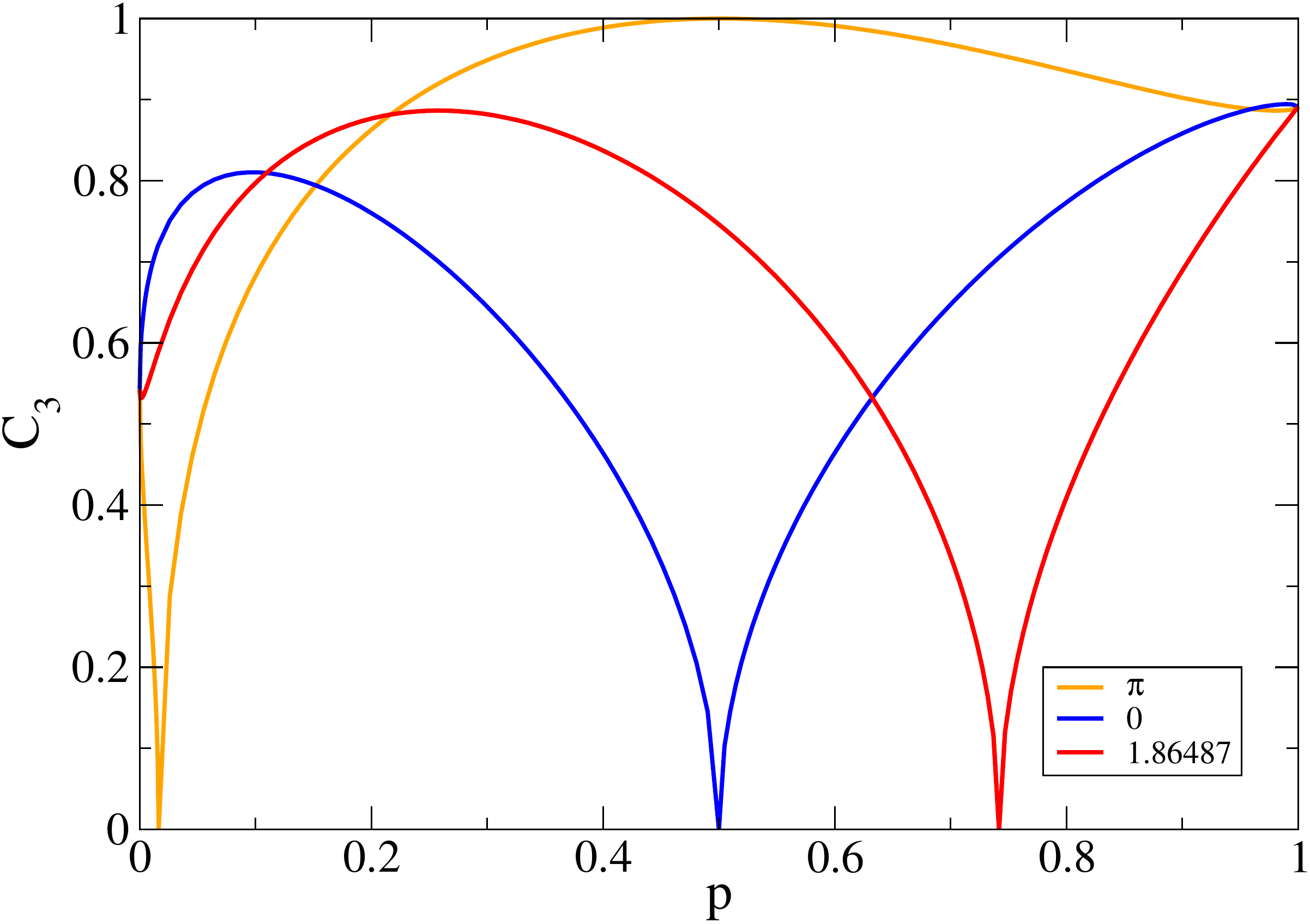}
  \caption{I show here the four characteristic curves for $C_3$ 
whose three-tangle becomes zero: two single real zeros at 
$p\approx 0.01636$ (solid orange curve) and 
$p=0.5$ (dashed blue curve) corresponding to an angle $\varphi=\pi$ and $\varphi=0$, respectively,
and the 
two coinciding curves which are zero at $p\approx 0.7418$ (dash-dash-dotted red curve).
The latest curve corresponds to two complex conjugate solutions $z_0$.
Both curves are for the angle $\varphi=\pm 1.8649=\arg(z_0)$.
The angles of $z$ for the different curves are shown in the legend.}
  \label{Char-Curves}
\end{figure}

We define the $n$-qubit GHZ- and W-states as
\beqa
\ket{GHZ_n}&=&\frac{1}{\sqrt{2}}(\ket{00\dots0}+\ket{11\dots1})\\
\ket{W_n}&=&\frac{1}{\sqrt{3}}(\ket{0\dots01}+\ket{0\dots10}+\dots\nonumber \\
&& \qquad +\ket{10\dots0}) 
\eeqa
where we consider the three-qubit example first
\beqa
\ket{GHZ_3}&=&\frac{1}{\sqrt{2}}(\ket{000}+\ket{111})\\
\ket{W_3}&=&\frac{1}{\sqrt{3}}(\ket{001}+\ket{010}+\ket{100})
\eeqa
and the density matrix
\beq
\rho(p)=p\ketbra{\psi_+}+(1-p) \ketbra{\psi_-}\; ,
\eeq
where
\beq
\ket{\psi_\pm}=\frac{1}{\sqrt{2}}(\ket{GHZ_3}\pm\ket{W_3})\; .
\eeq
These states satisfy the orthogonality condition $\braket{\psi_+}{\psi_-}=0$.
In order to calculate or estimate the three-tangle in $\rho(p)$,
we have to consider the characteristic curves\cite{LOSU,KENNLINIE}, hence
\beq
C_3(p,\varphi):=C_3(Z(p,\varphi))
\eeq
for the states
\beq
\ket{Z(p,\varphi)}:=\sqrt{p}\ket{\psi_+}-e^{i\varphi}\sqrt{1-p}\ket{\psi_-}\; .
\eeq
Some of them are shown in Fig.~\ref{Char-Curves} (more can be found in Ref.~\cite{JungPark16}).
It is hence useful to look for solutions $z_0$ to the equation
\beq\label{zet}
\tau_3(\ket{\psi_+}-z\ket{\psi_-})=0\; .
\eeq
The zeros $z_{0;j}$, $j=1,\dots,2n$ with $n\in\NN$, describe the vertices
of a zero-polytope, which becomes a three dimensionsional zero-simplex in this case.
I want to emphasize that the zero-simplex is an exact result 
and therefore the values $p$ of $\rho(p)$
which are lying inside the zero simplex are the only values 
for which the convex roof of $\rho(p)$ vanishes. Hence, it is also clear that 
the complement is made out of states with non-zero convex-roof.
The zeros of Eq.~\ref{zet} are 
\beqa
\fat{z_0}&=&(z_{0;1},z_{0;2},z_{0,3},z_{0,4})\\
&\approx& (1,-7.7543,0.5899 e^{1.8649 i},0.5899 e^{-1.8649 i}).
\eeqa
I want to emphasize that although the values for the zeros are exact, they are nevertheless 
approximated here since it is cumbersome to write them down analytically; in addition, 
I don't attribute to the knowledge of the exact values any further insight.
With $p_0=p(z_0)=1/(1+|z_0|^2)$, hence
\beqa
\fat{p_0}&=&(p_{0;1},p_{0;2},p_{0;3},p_{0;4})\nonumber\\
&\approx&(1/2,0.01636,0.74182,0.74182)\; ,
\eeqa
the values $p_0z_0$ are those to be 
convexely combined to zero\cite{OstUpperBound16,OstNineClasses16}.
The result is that for $p\in [0.11423,0.69289]$ the convex roof of
the three-tangle is zero.
The decomposition of $\rho(p)$ in $p=0.11423$ is given by
$\ket{Z(p_{0;1},0)}$ with weight $0.202362$ and  
$\ket{Z(p_{0;2},\pi)}$ with weight $0.797638$;
at $p=0.692885$ it is given by 
$\ket{Z(p_{0;1},0)}$ with weight $0.202362$ and the states  
$\ket{Z(p_{0;3}=p_{0;4},\pm 1.86487)}$ with weights $0.398819$ each. 
It is a curious coincidence that the weight of $\ket{Z(p_{0;1},0)}$ takes about the same value;
they deviate only by $3\times 10^{-16}$.

An upper bound to the convex-roof $\widehat{C_3}$ 
is shown in Fig.~\ref{fig2} together with the characteristic (gray background) curves:
\begin{figure}
\centering
\includegraphics[width=.9\linewidth]{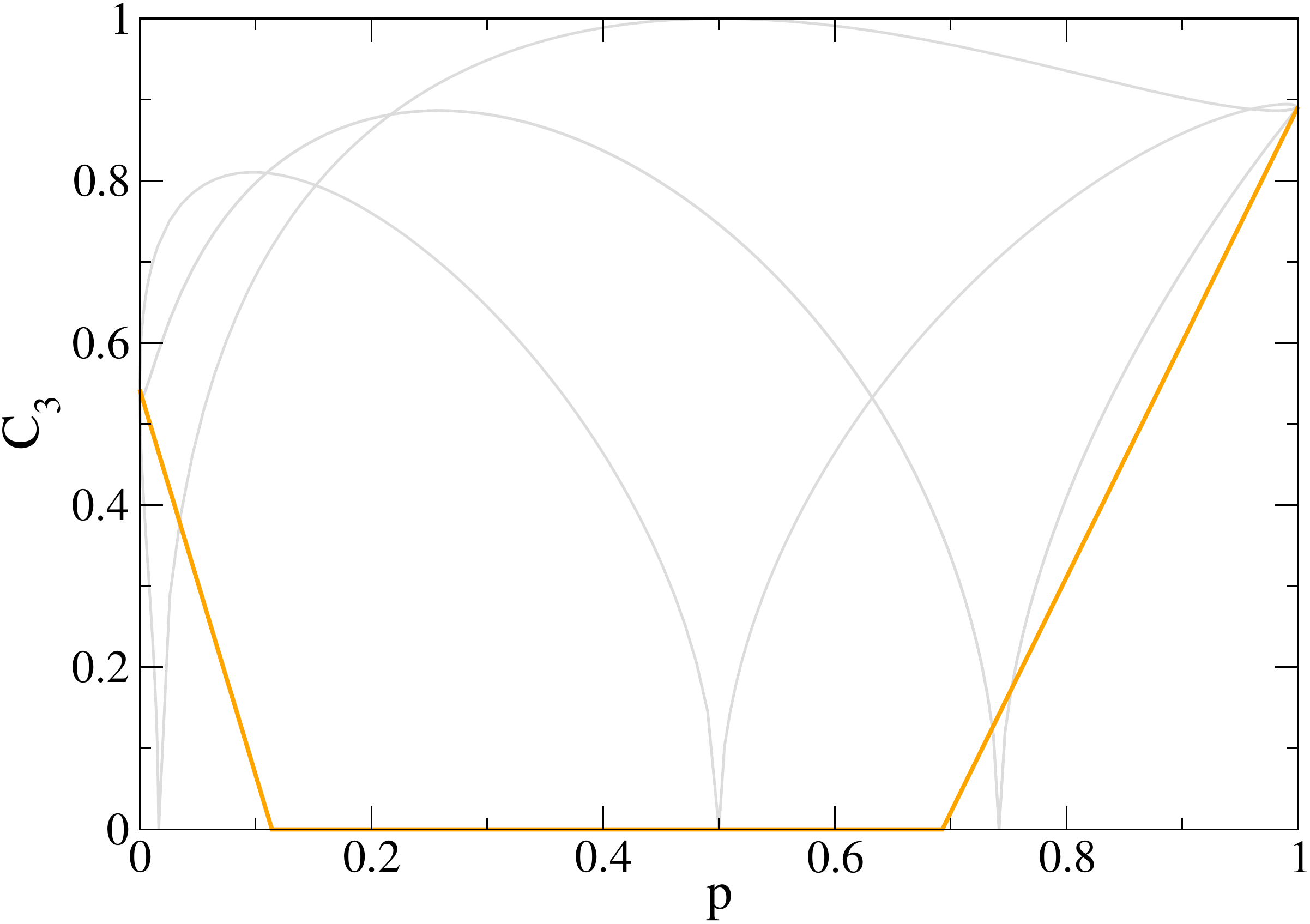}
  \caption{An upper bound to the convex-roof is shown for $\rho(p)$ (orange line).
It is piecewise linearly interpolating between $(p,C_3)=(0,\sqrt{8\sqrt{6}-9}/6)$,
$(0.11423,0)$, $(0.69289,0)$, and $(1,\sqrt{8\sqrt{6}+9}/6)$.
The intersection of $\rho(p)$ with the zero-simplex of the three-tangle is an exact result, 
whereas the linear extrapolation is certainly an upper bound to $C_3=|\sqrt{\tau_3}|$; 
it results from a superposition of the corresponding pure state and the density matrix 
with zero three-tangle closest to it.
Therefore the density matrix would be decomposed into three states for $0<p<0.114230$ and 
into four states for $0.692885<p<1$.
The characteristic curves are the gray curves in the background; they serve in order to demonstrate 
how the intersection with the zero-simplex, due to its convexity, 
leads to a shrinking of the region where $C_3[\rho(p)]=0$.}
  \label{fig2}
\end{figure}
the upper bound to the convex-roof is a piecewise straight (orange) line. 
I will therefore call it the {\em linearized} upper bound.
\subsection{Beyond linearization}\label{beyond:0}
The strong concavity of the characteristic curves
around their zeros, together with the fact that the plotted characteristic curves 
close to their zeros are a lower bound to other characteristic curves,
tells that whatever decomposition vector of the density matrix one will take
it will yield in a concave result at least in the vicinity of the zero-simplex.
This modifies close to $p=0$ or $p=1$ where it is rather likely that a piecewise convex 
curve might be obtained,
in particular in the interval $[0,0.11423]$ where one of the characteristic curves is
strongly convexly decreasing with a zero at $p\approx 0.01636$.
I therefore try for a slightly different decomposition here in order to check whether 
the convexity of this characteristic curve might
lead to a curve which somewhere lies below the straight line.
\begin{figure}
\centering
\includegraphics[width=.9\linewidth]{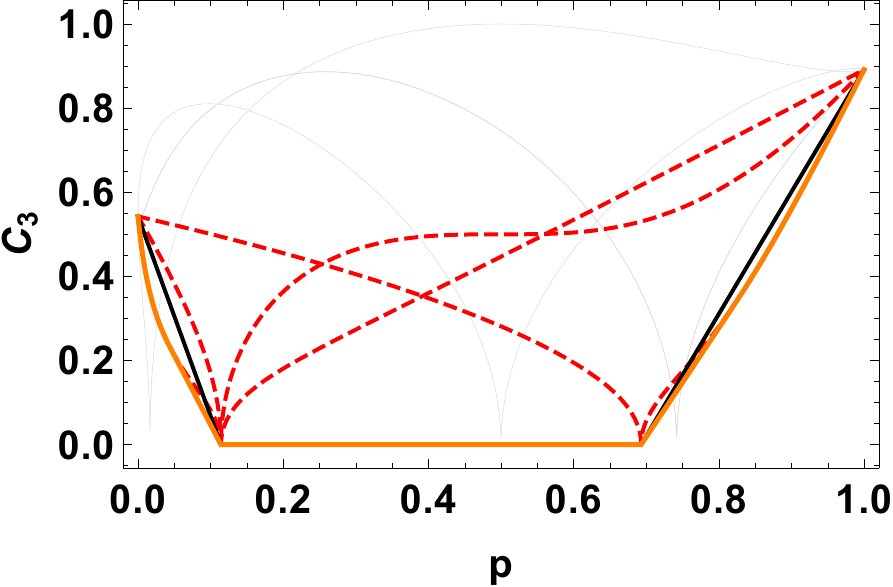}
  \caption{Here, I show results for some particular decompositions of $\rho(p)$
(read the text for details). The characteristic curve with a single zero at $p=0.01636$,
corresponding to an angle $\varphi=\pi$,
initially is strictly convex. 
Therefore that decompositions containing a state $\ket{Z(q,\pi)}$
will be also strictly convex close to the points $p=0,1$. 
That this is indeed the case also for $p$ close to $1$ is shown by the red dashed curves,
which comes to lie below the upper linearized bound (thin black line).
The corresponding new lower bound is the thick orange line.}
  \label{fig3}
\end{figure}
I chose to decompose the matrix $\rho(p)$ into two states, 
namely into the state $\ket{Z(p_{0;1},0)}$ and
the corresponding state $\ket{Z(q(p,p_{0;1}),\pi)}$ with $q(p,p_{0;1})$ 
in the interval given by $p$ and $p_{0;1}$ such that the 
line connecting the states
$\ket{Z(p_{0;1},0)}$ and $\ket{Z(q(p,p_{0;1}),\pi)}$ on the Bloch sphere hits the point
on the z-axis corresponding to $\rho(p)$. A further decomposition I had a look at,
is the equal mixture of the two states $\ket{Z(p_{0;3}=p_{0;4},\pm 1.86487)}$ 
with $\ket{Z(q(p,p_{0;1}),0)}$ such that the line interconnecting the two states is
again passing through $\rho(p)$.
The result is shown as red dashed lines in Fig.~\ref{fig3}. Some of them are lying 
below the straight line,
demonstrating that a better upper bound than the linearized one is obtained for the convex-roof 
$\widehat{C}_3$.
It is linear close to the borders of the interval $[0.11423,0.692885]$ 
up to $p_r=0.8240$ and down to $p_l=0.04395$, showing that the decomposition is made of 
convex decompostions of the two states $\ket{Z(p_{0;3}=p_{0;4},\pm 1.86487)}$ and a third state
$\ket{Z(q(p_{l/r},p_{0;1}),0)}$ (see Refs.~\cite{LOSU,KENNLINIE}). 
Beyond, it is strictly convex, telling that the decomposition
is here made of the two states $\ket{Z(p_{0;3}=p_{0;4},\pm 1.86487)}$ and the state
$\ket{Z(q(p,p_{0;1}),0)}$, which depends on $p$.

This procedure will be repeated for the general rank-two case in the next section.
It can be applied for general rank-two density matrices
and, using the results of Ref.~\cite{OstUpperBound16}, also for obtaining useful upper bounds
for general rank. It is a purely geometric method an therefore, it is not restricted to qubits.

\section{The interesting case}\label{general}

In order to demonstrate how the combined method of geometric view on the zero-polytope
with generalized decompositions to eventually going beyond the linearized method of Ref.~\cite{OstUpperBound16} 
works for the general case, we present the slightly modified example from Ref.~\cite{JungPark16}. 
\begin{figure}
\centering
\includegraphics[width=.9\linewidth]{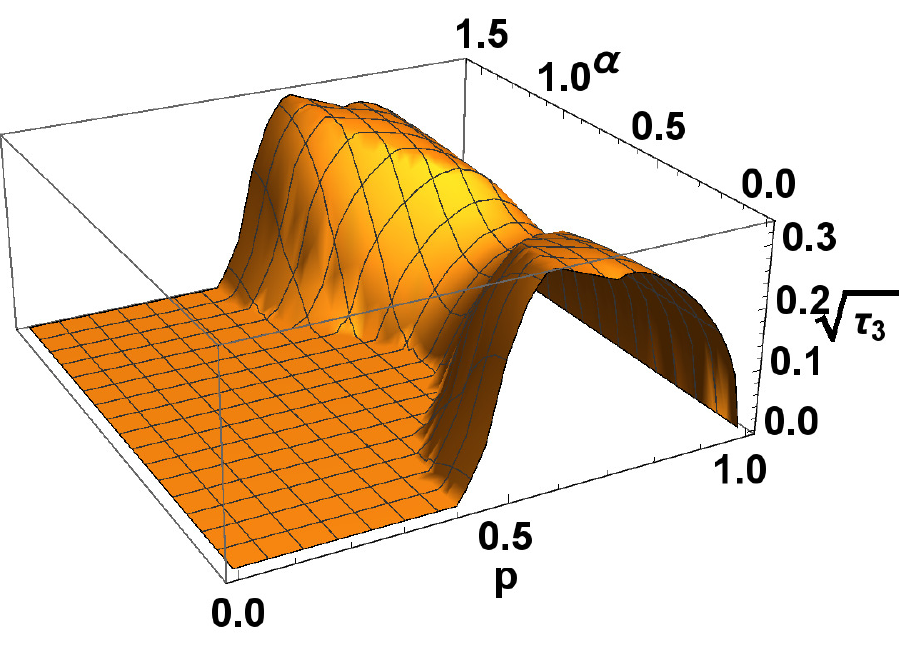}
  \caption{The upper bound for $\widehat{C_3}$ where one linearizes between the 
values for the states $\ket{\psi_i(p,\varphi)}$ and the corresponding extreme intersection 
points $p_{0:i}$, for $i=1,2$, of the line represented by $\rho(p)$ and the zero-simplex.}
  \label{sqrttau3}
\end{figure}

\subsection{The geometric view}
Thus, we turn to the more general example where the pure state
\beq
\ket{\Psi_4(p,\varphi)}:=\sqrt{p}\, \ket{GHZ_4} - \sqrt{1-p}e^{i\varphi} \, \ket{W_4}
\eeq
of four-qubits was given\cite{JungPark16}. It is a permutation invariant state
whose three-qubit density matrices, for their permutational symmetrie, all have the same form
\beqa
\rho_3(p,\varphi)&=&q(p) \ketbra{\psi_1(p,\varphi)} \nonumber\\
&&+ (1-q(p)) \ketbra{\psi_2(p,\varphi)}
\eeqa
with $q(p)=\frac{2+\sqrt{1-p^2}}{4}$ and
\beqa
\psi_1(p,\varphi)&=&f_1(p) e^{i\varphi}\ket{111}+g_1(p)\ket{000} \nonumber\\
&&\qquad+ h_1(p)e^{-i\varphi} \ket{W_3}\\
\psi_2(p,\varphi)&=&f_2(p) e^{i\varphi}\ket{111}+g_2(p)\ket{000}  \nonumber\\
&&\qquad+ h_2(p)e^{-i\varphi} \ket{W_3}\; .
\eeqa
Here, the functions are defined as
\beqa
f_1(p)&:=&\sqrt{\frac{2}{(1+p)(3-p)+(3+p)\sqrt{1-p^2}}}\,p\\
g_1(p)&:=&\sqrt{p\frac{4\sqrt{1-p^2}-3p+5}{(3+p)\sqrt{1-p^2}+(1+p)(3-p)}}\\
h_1(p)&:=&\sqrt{\frac{3p(1-p)}{(1+p)^2-(1-p)\sqrt{1-p^2}}}\\
f_2(p)&:=&\sqrt{\frac{2}{(1+p)(3-p)-(3+p)\sqrt{1-p^2}}}\, p\\
g_2(p)&:=&\sqrt{p\frac{4\sqrt{1-p^2}+3p-5}{(3+p)\sqrt{1-p^2}-(1+p)(3-p)}}\nonumber\\
&&\sign{(3-5p)}\\
h_2(p)&:=&-\sqrt{\frac{3p(1-p)}{(1+p)^2+(1-p)\sqrt{1-p^2}}}\; .
\eeqa
The three-tangle is a periodic function of $\varphi$ with period $\pi/2$,
because of the four qubit permutation symmetry of the state. 
We show the results of the algorithm from Ref.~\cite{OstUpperBound16},
which except the default linearization gives an exact result for the zeros, in Fig.~\ref{sqrttau3}.
It is an upper bound to $\widehat{C_3}$.
\begin{figure}
\centering
\includegraphics[width=.47\linewidth]{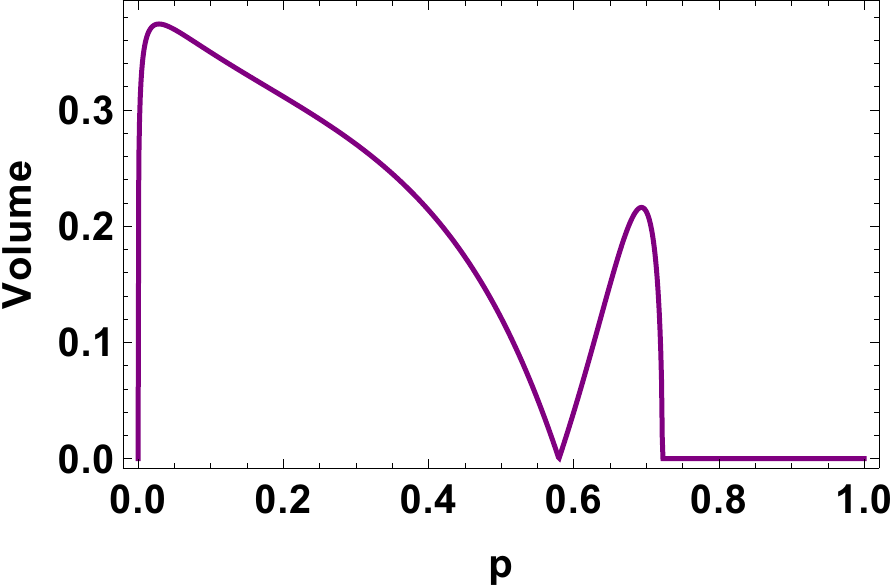}
\includegraphics[width=.47\linewidth]{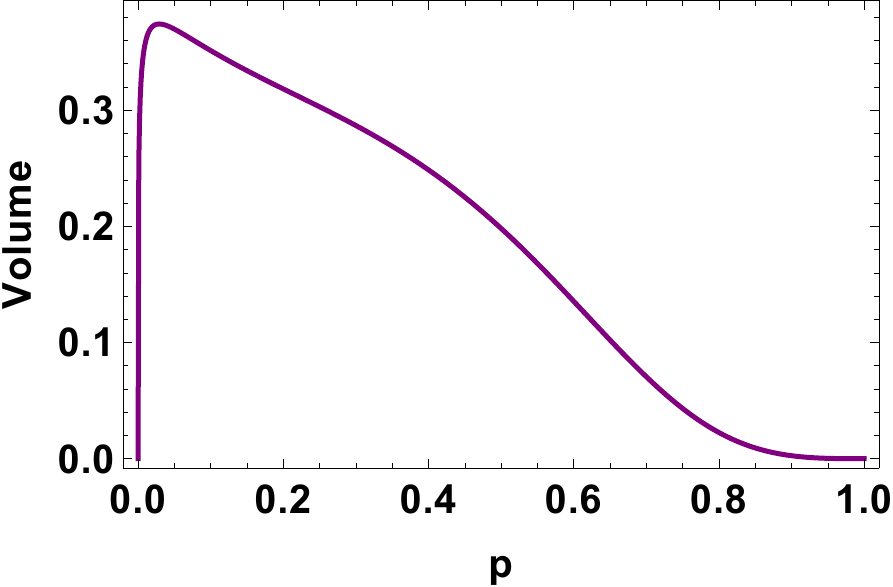}
  \caption{The volume of the zero-simplex for two values of $\varphi=0$ (left panel)
and $\varphi=\pi/4$ (right panel). For $\varphi=0$ the volume grows to a finite value for diminishing again
unless it is crossing with zero volume (staying however two-dimensional) to grow again up to 
a value of $p=0.722074$ where it again becomes two-dimensional up to $p=1$. 
Here, the imaginary part of the two corresponding solutions is zero and we have again four 
real values. This passage through zero
in between is missing for $\varphi=\pi/4$; in particular the zero-simplex is always 
three-dimensional for $p\in \,(0,1)$.}
  \label{Vol}
\end{figure}

\subsection{Beyond linearization}
In order to test whether it is possible also here to come below the linearized upper bound, 
I checked the zeros of Eq.~\ref{zet} and 
the particular decompositions I have described in detail in the last section.\\
In $[0.722074,1]$, there are 4 real solutions.
For the remaining values of $p$,
there are two complex conjugate solutions besides two which stay real.
One decomposition for which the three-tangle vanishes is always made from real solutions here, 
whereas the other one is made 
out of three pure states: one corresponding to a real solution 
and the two complex conjugate solutions.
The zero-simplex is varying its dimension as shown in Fig.~\ref{Vol} 
for $\varphi=0$ and $\varphi=\pi/4$ respectively.
\begin{figure}
\centering
\includegraphics[width=.9\linewidth]{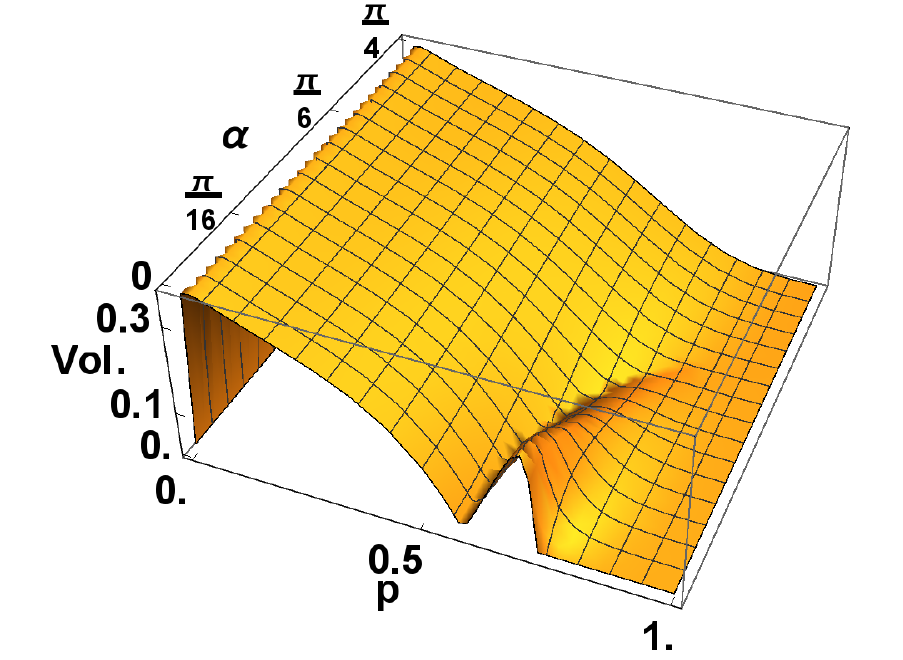}
  \caption{A three-dimensional plot of the zero-simplex dimension.}
  \label{Vol-3d}
\end{figure}
It is becoming zero twice for $\varphi=0$: a single point, 
where the line spanned by the 
complex conjugate values with non-zero imaginary part crosses
the corresponding line between the two other real values,
and there is a whole interval $[0.722074,1]$ for $p$ where the zero-simplex 
is two-dimensional. There, four real solutions appear.
This feature however is not stable against small perturbations in $\varphi$.\\
The single zero disappears for $\varphi\gtrsim 0.5236$
with the zero-simplex being everywhere three-dimensional (except at the boundaries);
in particular for $\varphi=\pi/4$.
This is indicated in Fig.~\ref{Vol-3d}.

\begin{figure}
\centering
\includegraphics[width=.9\linewidth]{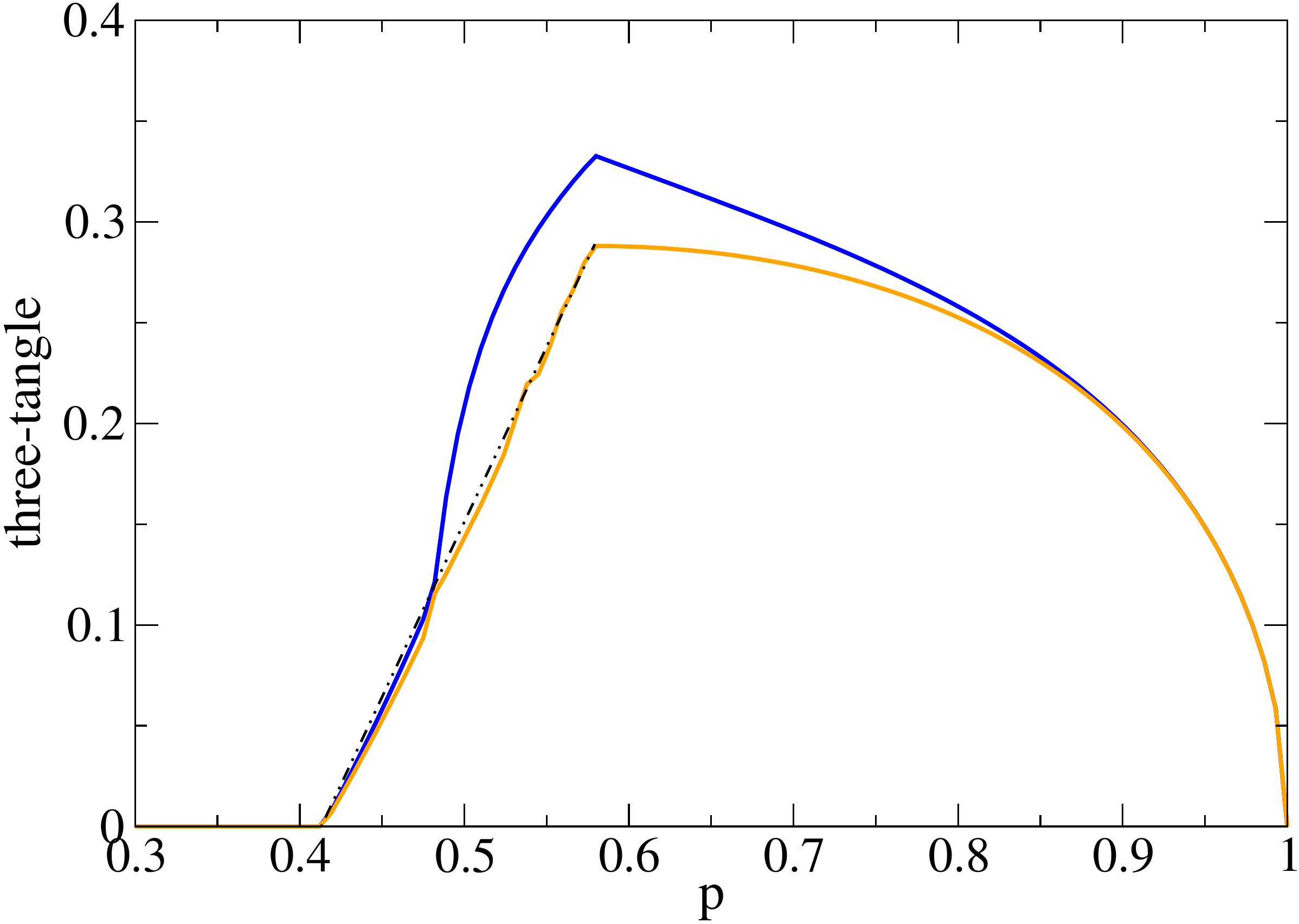}
  \caption{Two upper bounds for $\widehat{C_3}$ for $\varphi=0$ as a function of $p$.
Besides the linearized version from Ref.~\cite{OstUpperBound16} (upper blue curve) also 
the one coming out of the procedure described here (see discussion of Fig.~\ref{fig3}) is shown
(orange lower curve). This curve is well approximated with the straight black dash-dotted 
line in the figure. 
It can be seen however that the convex-roof lies at least slightly below the straight line.}
  \label{sqrttau3-ext}
\end{figure}
An upper bound to the three-tangle $\widehat{C_3}$ is shown in Fig.~\ref{sqrttau3-ext} for $\varphi=0$ 
in the linearized version and the procedure described in 
section ~\ref{beyond:0} (see also the discussion of Fig.~\ref{fig3} in the text). 
It is seen that both basically coincide close to the zeros but they 
deviate considerably in between. 
This is not the case for $\varphi=\pi/4$, where both curves coincide (not shown here). 

\section{Extended monogamy}\label{monogamy}

It is clear that the residual tangle is not measured in general 
by an SL-invariant quantity\cite{NoMonogamy}.
\begin{figure}
\centering
\includegraphics[width=.95\linewidth]{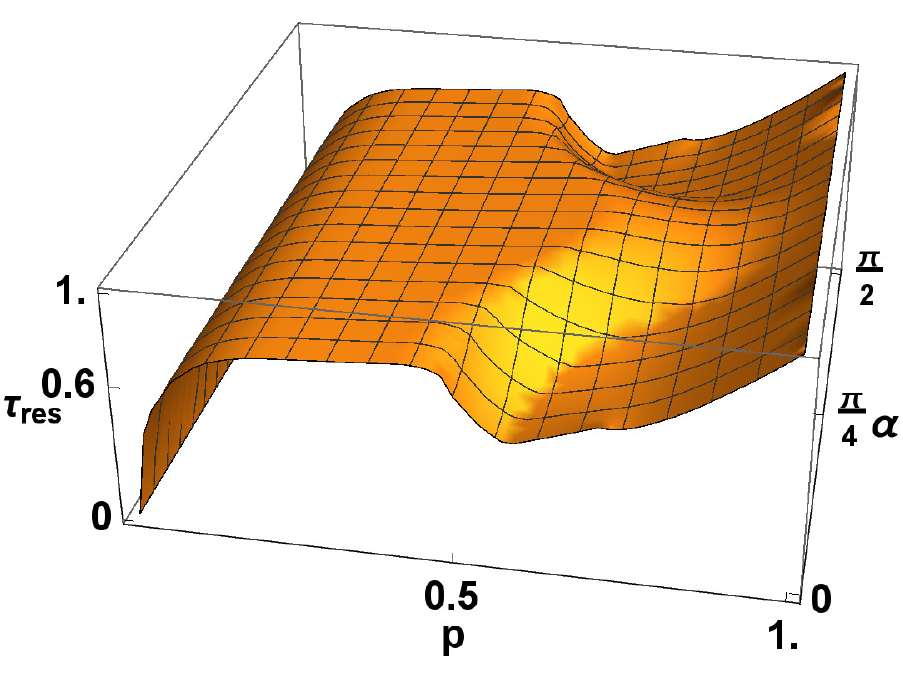}
  \caption{The extended residual tangle\cite{AdessoRegula14,AdessoRegula14b,AdessoRegula16,AdessoOsterlohRegula16} 
using $\widehat{C_3}^2$ as the measure for the three-tangle.
I do not show the outcome for $\widehat{\sqrt{C_3}}^4$, since it is smaller than 
$\widehat{C_3}^2$\cite{OstNineClasses16} and accordingly the residual tangle is bigger. }
  \label{fig:monogamy}
\end{figure}
Therefore it makes little sense to subtract from the residual tangle which has no SL-invariance 
an SL-invariant quantity. When nevertheless doing so, one recognizes that the monogamy cannot be extended 
with the usual three-tangle $\widehat{\tau_3}$ or even its square root 
$\widehat{\sqrt{\tau_3}}^2=\widehat{C_3}^2$\cite{AdessoRegula14,AdessoRegula14b,AdessoRegula16}. 
The ultimate possibility would be $\widehat{\sqrt[4]{\tau_3}}^4$, 
which could not be excluded for pure states of four qubits\cite{AdessoOsterlohRegula16}.
This doesn't mean that it won't be excluded for some $n$-qubit pure state with $n>4$.
This question has to be answered in future work.
As far as the extended monogamy relations are concerned, the states already satisfy it taking
$\widehat{C_3}^2$ as measure for the three-tangle.
This can be seen in Fig.~\ref{fig:monogamy} taking the linearized upper bound for $\widehat{C_3}^2$; 
it therefore provides a lower bound for the residual 'four-tangle'. It is ranging from zero 
(for the W-states) to one (for the GHZ-states).

\section{Conclusions} \label{concls}

I have presented a method how intersections of a certain density matrix of rank two with 
the zero-polytope can be calculated exactly. This is an exact solution to any problem of non-coinciding 
zeros of the zero-polytope, as inserted in the algorithm of Ref.~\cite{OstUpperBound16}.
I have exemplified this method on an open problem recently raised by Jung and Park\cite{JungPark16}.
I have described in detail for the toy example of Ref.~\cite{JungPark16} how the simplest 
linearized version of an upper bound can be obtained, and how one can go beyond it. 
To this end, I calculate a meaningful upper bound of the three-tangle $\sqrt{\tau_3}$ for their toy-example
which is better than the linear interpolation in Ref.~\cite{OstUpperBound16}.
As a proof of principles, I apply this formalism further to the general case of 
superpositions of four-particle GHZ and W states,
calculating the linearized form for the upper bound together with the extended version 
for $\sqrt{\tau_3}$.
As a byproduct I briefly comment on the extended CKW-monogamy and provide a graph also for
a generalized 'four-tangle'.
I want to mention that the calculation of the three-tangle of 
$\rho=p \ketbra{GHZ_4}+(1-p) \ketbra{W_4}$ is trivially zero 
for each three-qubit subsystem. 

As purely geometrical procedure the findings of this work are 
applicable to obtaining the zeros of general SL- and also
of arbitrary SU-invariant polynomial entanglement measures
with bidegree $(d_1,d_2)$\cite{Luque06,JohanssonO13}; this holds as well for the procedure
of going beyond the linear interpolation. 
They are also applicable to qudits.
The same line of thoughts 
can be adopted to arbitrary rank density matrices\cite{OstUpperBound16}.\\

\section*{Acknowledgements}

I acknowledge discussions with K. Krutitsky and R. Sch\"utzhold.


\end{document}